\documentclass[aps,prl,reprint,superscriptaddress,preprintnumbers,amsmath,amssymb]{revtex4-2}
\usepackage{graphicx}
\usepackage{dcolumn}
\usepackage{bm}

\begin{document}


\title{
  Cohesion-induced hysteresis and breakdown of marginal stability in jammed granular materials
}

\author{Michio Otsuki}
 \email{otsuki@riko.shimane-u.ac.jp}
\affiliation{%
    Institute of Science and Engineering, Shimane University, Shimane 690--8504, Japan
}
\author{Kiwamu Yoshii}%
\affiliation{%
    Department of Applied Physics, Tokyo University of Science, Tokyo 125--8585, Japan
}%
\author{Hideyuki Mizuno}
\affiliation{%
Graduate School of Arts and Sciences, The University of Tokyo, Tokyo, 153--8902, Japan
}%

\date{\today}

\begin{abstract}
  The dependence of mechanical properties on microscopic interactions remains a central problem in the physics of disordered solids near the jamming transition.
We numerically and theoretically investigate the mechanical response of jammed cohesive granular materials using discrete element simulations and effective medium theory (EMT).
We find that the shear modulus exhibits pronounced hysteresis under compression and decompression, even though the interparticle force law itself is strictly history-independent.
While such hysteresis disappears for purely repulsive particles when mechanical properties are characterized in terms of pressure, it persists in cohesive packings, indicating that pressure is not a unique state variable for cohesive particles.
Extending EMT to cohesive interactions, we show that the functional form of the shear modulus remains the same for both repulsive and cohesive particles, but that attractive interactions violate marginal stability.
The resulting deviation from marginal stability generates excess rigidity, as predicted by a scaling relation.
This prediction is quantitatively verified by numerical simulations and explains the persistent hysteresis in cohesive packings.
\end{abstract}

\maketitle



{\it Introduction.---}
Disordered packings of particles, such as granular materials, colloidal suspensions, emulsions, and foams, acquire rigidity when the packing fraction $\phi$ exceeds a critical value $\phi_{\rm J}$ \cite{Liu1998,Hecke2010,Behringer2019}.
This transition to a solid-like state is known as the jamming transition.
Near $\phi_{\rm J}$, critical behaviors emerge: the vibrational density of states (vDOS) exhibits a plateau below a characteristic frequency, and mechanical properties such as the shear modulus $G$ follow power-law scaling with the distance from jamming, $\phi - \phi_{\rm J}$~\cite{OHern2002,OHern2003,Wyart2005a,Wyart2005b,Otsuki2017}.
However, it has been established that the value of $\phi_{\rm J}$ is not unique but depends on the preparation protocol of the packing, including annealing, compression, and shear training \cite{Chaudhuri2010,Kumar2016,Vagberg2011,Ozawa2012,Ozawa2017,Urbani2017,Jin2021,Kawasaki2024}.
Such protocol dependence, related to shear jamming \cite{Bi2011,Vinutha2016,Nagasawa2019,Otsuki2020,Babu2021,Pan2023a,Pan2023b}, implies that mechanical properties measured at a fixed packing fraction $\phi$ exhibit pronounced history dependence \cite{Chaudhuri2010,Kumar2016}.


This protocol dependence motivates the use of alternative state variables to characterize mechanical properties near jamming, such as the pressure $p$ and coordination number $Z$ \cite{Chaudhuri2010,Zheng2018,Jin2021,Kawasaki2024}.
Previous theoretical studies express $G$ in terms of $p$ or $Z$~\cite{Wyart2005a,Zaccone2011,Tighe2011,Wyart2010,Yoshino2014,DeGiuli2014a,DeGiuli2014b,Mao2010,Mizuno2024}.
In particular, effective medium theory (EMT) expresses $G=G(p,Z)$ for elastic networks near jamming~\cite{DeGiuli2014a,DeGiuli2014b,Mao2010,Mizuno2024}.
For assemblies of purely repulsive particles, jammed states are known to be marginally stable~\cite{Wyart2005a,Wyart2005b,Ikeda2022}, which establishes a relation between $p$ and $Z$.
Under this marginal stability condition, EMT reduces to a single-variable scaling, yielding $G\sim Z-Z_{\rm iso}\sim p^{1/2}$ for linear repulsive interactions~\cite{OHern2002,OHern2003}, where $Z_{\rm iso}$ is the isostatic coordination number~\cite{Hecke2010}.
Taken together, these results suggest that the apparent history dependence is largely eliminated when mechanical properties are characterized by $p$ or $Z$ rather than $\phi-\phi_{\rm J}$~\cite{Goodrich2014,Pan2023a}.

However, most theoretical frameworks of jamming have been developed for purely repulsive particles, while attractive interactions, such as capillary and wetting forces in granular materials \cite{Mitarai2006,Herminghaus2023}, are ubiquitous in realistic systems. 
It therefore remains unclear whether the same description applies to cohesive systems. 
Recent numerical studies have indicated that cohesive interactions significantly modify particle configurations and mechanical responses near jamming \cite{Head2007,Chaudhuri2012,Irani2014,Irani2016,Katgert2013,Koeze2018,Koeze2020,Lois2008,Yoshii2025,Zheng2016}, yet a unified theoretical framework and a systematic characterization of their protocol dependence are still lacking.
In this Letter, we investigate cohesive granular packings using
$\phi$-controlled compression and decompression simulations.
We show that cohesive packings retain intrinsic history dependence even when mechanical properties are expressed in terms of pressure, in contrast to purely repulsive systems.
We explain this behavior by extending EMT to
cohesive particles and demonstrating the breakdown of marginal
stability.

\begin{figure*}[htbp]
  \begin{center}
       \includegraphics[width=1.0\linewidth]{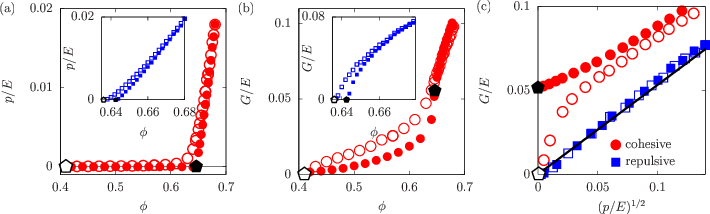}
  \end{center}
    \caption{
  (a) Pressure $p$ as a function of the packing fraction $\phi$ for
cohesive particles (main panel) and repulsive particles (inset).
(b) Shear modulus $G$ as a function of $\phi$ for cohesive particles
(main panel) and repulsive particles (inset).
(c) Shear modulus $G$ as a function of $p^{1/2}$ for cohesive and
repulsive particles.
Compression and decompression data are shown by open and closed
symbols, respectively.
The packing fraction $\phi_{p=0}$ where the pressure vanishes is
indicated by black pentagons.
The solid line in (c) indicates the expected scaling
$G \propto p^{1/2}$ for repulsive particles.
    }
    \label{P_G}
\end{figure*}

{\it Methods and protocol.---}
We consider a system of frictionless monodisperse particles in a cubic box of size $L$ with periodic boundary conditions.  The interparticle force is defined as a piecewise linear function of the interparticle distance $r$, with a repulsive core, a short-range attractive tail, and a cutoff distance $\ell_{\rm c}$
~\cite{Irani2014,Irani2016,Koeze2018,Koeze2020}:
\begin{align}
f(r) =
\begin{cases}
  k \left[ (1 - 2\alpha)\,\ell_{\rm c} - r \right]
  & r < (1-\alpha)\ell_{\rm c}, \\
 -k \left( \ell_{\rm c} - r \right)
  & (1-\alpha)\ell_{\rm c} \le r < \ell_{\rm c}, \\
  0 & r \ge \ell_{\rm c}.
\end{cases}
\label{eq:force}
\end{align}
Here, $k$ is the elastic constant and $\alpha$ is the dimensionless attraction strength, controlling the maximum attractive force $\alpha k \ell_{\rm c}$ at $r=(1-\alpha)\ell_{\rm c}$.  
The resulting shape of the interparticle force is shown later.
It is a simplified variant of models for wet granular materials~\cite{Roy2016} and exhibits no microscopic interaction hysteresis.
The case $\alpha=0$ corresponds to a purely repulsive linear interaction \cite{OHern2002,OHern2003}.  
Different conventions for length and interaction strength in previous studies~\cite{Koeze2018,Koeze2020} are trivially related to the present one, as detailed in the Supplemental Material~\cite{Supple}.


We vary the packing fraction $\phi$ by controlling the system size $L$
and integrate particle dynamics using the discrete element method
(DEM)~\cite{Cundall1979}, where a velocity-dependent dissipative force
is included.
The initial state is prepared at a small packing fraction $\phi_{\rm I}$,
where particles are randomly placed without overlap.
The packing fraction is then increased stepwise by $\Delta\phi$
until it reaches a maximum value $\phi_{\rm max}$; we refer to this
process as compression.
Subsequently, $\phi$ is decreased stepwise by $\Delta\phi$,
which we refer to as decompression.
At each $\phi$ during both compression and decompression,
the pressure $p$ and shear modulus $G$ are measured after the system
reaches mechanical equilibrium.
We refer to $\alpha=0$ as {\it repulsive} and to $\alpha=0.001 > 0$ as {\it cohesive}.
Stress and time are nondimensionalized using $E \equiv k/\ell_{\rm c}$ and $\tau \equiv \sqrt{m/k}$, respectively.
Further details of the simulation protocol, as well as results for
various values of $\alpha$ obtained using pressure-controlled protocols,
are provided in the Supplemental Material~\cite{Supple}.

{\it Mechanical hysteresis.---}
Figure~\ref{P_G}(a) shows the pressure $p$ as a function of the packing fraction $\phi$ for cohesive particles (main panel) and repulsive particles (inset).
Data obtained during compression and decompression are indicated by open and closed symbols, respectively; data points with zero pressure are omitted for clarity~\cite{Supple}.
We denote by $\phi_{p=0}$ the packing fraction where the pressure vanishes (pentagons in Fig.~\ref{P_G}).
During compression, $p$ becomes positive once $\phi$ exceeds
$\phi_{p=0}$, although for cohesive particles its magnitude remains small over a broad range of $\phi$, making the onset of
finite pressure visually subtle in the main panel.
During decompression, the pressure decreases to zero at a larger value of $\phi$, so that $\phi_{p=0}$ exhibits a clear protocol dependence, which is significantly more pronounced for cohesive particles than for repulsive ones.
For repulsive particles, $\phi_{p=0}$ corresponds to the jamming
point, and states with $|p|>0$ do not exist for $\phi<\phi_{p=0}$.
By contrast, cohesive particles exhibit states with $\phi<\phi_{p=0}$ during decompression, corresponding to negative pressure.
Alternative representations highlighting the low-$|p|$ regime are
provided in the Supplemental Material~\cite{Supple}.

The corresponding behavior of the shear modulus is shown in
Fig.~\ref{P_G}(b), where $G$ is plotted as a function of the packing
fraction $\phi$.
For repulsive particles, $G$ exhibits hysteresis between compression
and decompression at a given $\phi$, reflecting the protocol
dependence of $\phi_{p=0}$.
For cohesive particles, hysteresis is also observed, but with a
qualitatively different feature: during decompression, the shear
modulus remains finite even for $\phi<\phi_{p=0}$.
This implies $G>0$ at negative pressure ($p<0$), in contrast to the
repulsive case where rigidity disappears once $p$ vanishes.
The persistence of rigidity at $p<0$ originates from cohesive
interactions, which stabilize packings even in
the absence of compressive stresses.

Replotting the shear modulus as a function of $p^{1/2}$ for $p\ge0$,
as shown in Fig.~\ref{P_G}(c), highlights the scaling relation
$G\propto p^{1/2}$ established for repulsive particles
\cite{OHern2002,OHern2003}.
The behavior for $p<0$ is shown in the Supplemental
Material~\cite{Supple}.
The repulsive-particle data collapse onto a linear master curve,
indicating that pressure serves as an effective state variable for
repulsive packings near jamming.
In contrast, cohesive particles exhibit pronounced hysteresis:
during compression, $G$ increases from zero at $p=0$,
whereas during decompression, $G$ remains finite even as $p\to0$.
Thus, the scaling relation $G\propto p^{1/2}$ breaks down for cohesive
packings, and hysteresis persists even in the $G(p)$ representation.
The hysteresis observed in $G$ and $p$ therefore arises despite the
absence of microscopic interaction hysteresis and instead reflects the
evolution of mechanically stable contact networks.
This observation suggests that an additional structural variable,
beyond pressure, is required to characterize the mechanical state of
cohesive packings.

\begin{figure}[htbp]
  \centering
        \includegraphics[width=0.95\linewidth]{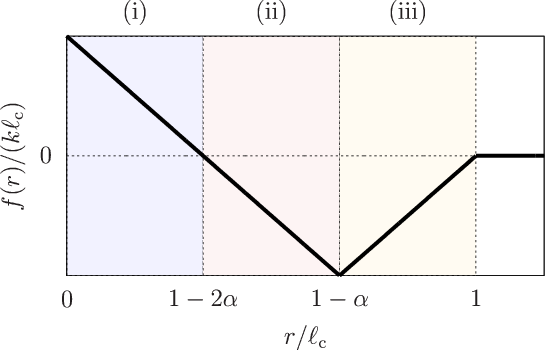}
  \caption{
Schematic of the static interparticle force $f(r)$ as a function of the interparticle distance $r$.
  The repulsive (i), stabilizing attractive (ii), and locally destabilizing (iii) regimes are indicated by different shaded regions.
  }
\label{Force}
\end{figure}

\begin{figure*}[htbp]
  \begin{center}
        \includegraphics[width=1.0\linewidth]{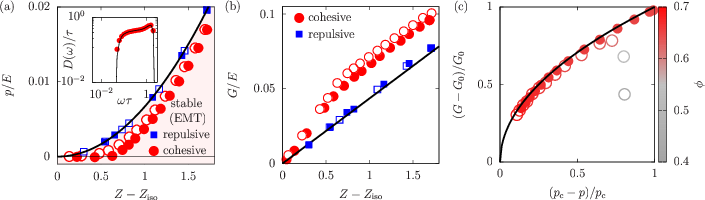}
\caption{ 
Marginal stability and its breakdown in repulsive ($\alpha=0$) and cohesive ($\alpha>0$) packings, compared with EMT predictions.
Open and closed symbols correspond to decompression and compression,
respectively.
(a) Pressure $p$ as a function of $Z-Z_{\rm iso}$.
The solid line indicates the marginal-stability condition
$p=p_{\rm c}(Z)$.
The shaded region indicates EMT-stable states satisfying
$p\le p_{\rm c}(Z)$.
(Inset) vDOS $D(\omega)$ for cohesive
particles at $\phi=0.60$ during compression.
The symbols and solid line represent the spectra obtained from the
original and unstressed Hessians, respectively.
(b) Shear modulus $G$ as a function of $Z-Z_{\rm iso}$.
The solid line represents the bare shear modulus
$G=G_0(Z)$ introduced in EMT.
(c) Scaling plot of the excess rigidity $(G-G_0)/G_0$
as a function of the deviation from marginal stability
$(p_{\rm c}-p)/p_{\rm c}$ for cohesive packings,
testing the EMT prediction.
The solid line represents the EMT prediction,
Eq.~\eqref{eq:general_scaling}.
The color of the symbols represents the packing fraction $\phi$.
    }
\label{3figures2}
  \end{center}
\end{figure*}

{\it Effective medium theory.---}
To explain this behavior, we employ EMT for cohesive packings~\cite{Mizuno2024}, where the mechanical response is governed by the Hessian $M$ constructed from interparticle forces.
To analyze the effect of cohesion, we examine the force law $f(r)$ in Eq.~\eqref{eq:force} (Fig.~\ref{Force}), which satisfies $f(r)=-U'(r)$ with a pair potential $U(r)$.
This force law consists of three regimes:
a repulsive regime with $f(r)\ge0$,
a stabilizing attractive regime with $f(r)<0$ and $f'(r)<0$,
and a locally destabilizing regime where $f(r)<0$ and $f'(r)>0$,
corresponding to a negative effective stiffness.
To quantify local stability in the Hessian, we define
$s(r)\equiv -f'(r)/k$, with $s=1$ for repulsive/stabilizing
and $s=-1$ for locally destabilizing contacts.
For purely repulsive particles ($\alpha=0$), the destabilizing regime is absent, so $s(r)=1$ for all pairs.
Thus, $s(r)$ indicates whether a contact stabilizes or destabilizes the Hessian.

To make this explicit, we represent the Hessian through the quadratic variation of the potential energy,
$\delta \mathcal{U}=\frac{1}{2}\langle u|M|u\rangle$,
for small displacements $\{\vec{u}_i\}$ with $|u\rangle=[\vec{u}_1,\ldots,\vec{u}_N]$:
\begin{equation}
M=\sum_{\langle ij\rangle} |ij\rangle
\left[
s_{ij}k\,\vec{n}_{ij}\otimes\vec{n}_{ij}
-\frac{f_{ij}}{r_{ij}}\left(I-\vec{n}_{ij}\otimes\vec{n}_{ij}\right)
\right]
\langle ij|.
\label{eq:M}
\end{equation}
Here, $\vec{n}_{ij}$ is the unit normal vector pointing from particle $j$ to $i$, $I$ is the $3 \times 3$ identity matrix, $r_{ij}$ is the distance between particles $i$ and $j$, $f_{ij}=f(r_{ij})$, and $s_{ij}=s(r_{ij})$.
The bra-ket notation selects relative displacements,
$\langle ij|u\rangle=\vec{u}_i-\vec{u}_j$.

For $\alpha=0$, one has $s_{ij}=1$ for all contacting pairs, and the
Hessian $M$ reduces to that of purely repulsive particles
analyzed in previous studies
\cite{Wyart2005a,Wyart2005b,DeGiuli2014a,DeGiuli2014b,Mizuno2024}.
For cohesive particles ($\alpha>0$), the force law formally allows contacts with $s_{ij}=-1$ corresponding to the locally destabilizing regime.
However, in the static configurations obtained from our DEM
simulations, we find that such contacts are absent
(see Supplemental Material~\cite{Supple}).
This indicates that contacts in the locally destabilizing regime are eliminated during mechanical relaxation because they are unstable against infinitesimal perturbations, so that effectively $s_{ij}=1$ for all contacts.
As a result, the form of the Hessian $M$ for cohesive packings coincides with that for repulsive particles.
We can therefore directly apply the EMT framework developed for
repulsive systems to cohesive packings.

Within EMT, the system described by the Hessian matrix in
Eq.~\eqref{eq:M} is represented by a random spring network that captures
the same local force structure.
The network provides a mean-field description of the original packing
characterized by the pressure $p$ and coordination number
$Z$~\cite{DeGiuli2014b,Mizuno2024}.
Because the Hessian $M$ is effectively identical
for repulsive and cohesive packings,
the EMT prediction for the shear modulus is identical
to that of purely repulsive systems \cite{DeGiuli2014b,Mizuno2024}:
\begin{equation}
G = G(Z,p) = G_0(Z)\left[1+\sqrt{1-\frac{p}{p_{\rm c}(Z)}}\right],
\label{eq:G}
\end{equation}
See Supplemental Material~\cite{Supple} for the derivation.
Here,
\begin{equation}
G_0(Z)=E_0\,(Z-Z_{\rm iso}),
\label{eq:G0}
\end{equation}
is the bare shear modulus without the correction due to the second term arising from the second term in Eq.~\eqref{eq:G}, and
\begin{equation}
p_{\rm c}(Z)=\Pi_0\,(Z-Z_{\rm iso})^2
\label{eq:pc}
\end{equation}
is the critical pressure, with prefactors $E_0$ and $\Pi_0$.
Notably, $G(Z,p)$, $G_0(Z)$, and $p_{\rm c}(Z)$ do not explicitly depend
on the attraction strength $\alpha$ and therefore apply to both
repulsive ($\alpha=0$) and cohesive ($\alpha>0$) systems within EMT.
For $p>p_{\rm c}(Z)$, the EMT expression yields no real solution for
$G$, indicating that mechanically stable configurations must satisfy
$p\le p_{\rm c}(Z)$~\cite{DeGiuli2014b,Mizuno2024}.

For purely repulsive particles, jammed configurations are known to exhibit marginal stability, characterized by the condition
$p=p_{\rm c}(Z)$.
Under marginal stability, using Eq.~\eqref{eq:pc}, the pressure $p$ and coordination number $Z$ are no longer independent but satisfy $Z-Z_{\rm iso}=(p/\Pi_0)^{1/2}$.
Substituting this relation into Eq.~\eqref{eq:G}, the shear modulus reduces to $G=G_0(Z)$ and obeys the scaling $G\propto Z-Z_{\rm iso}\propto p^{1/2}$, as expected for repulsive packings near jamming.
By contrast, for cohesive particles, marginal stability has not been established.
In this case, EMT only imposes the stability condition
$p\le p_{\rm c}(Z)$.
Consequently, the pressure $p$ and coordination number $Z$ cannot be reduced to a single state variable, and the shear modulus generally deviates from the bare value as $G\ge G_0(Z)$, explaining the persistent hysteresis observed in Fig.~\ref{P_G}(c).

{\it Breakdown of marginal stability.---}
To examine marginal stability and its predictions, we analyze the results of our DEM simulations in terms of the pressure $p$, coordination number $Z$, and shear modulus $G$.
Here, $Z$ denotes the average number of force-bearing contacts per particle, excluding rattlers, measured in mechanically stable configurations.
For comparison with numerical data, we identify $p_{\rm c}(Z)$ and $G_0(Z)$ with the relations $p(Z)$ and $G(Z)$ measured for purely repulsive particles ($\alpha=0$) during compression, where marginal stability implies $p=p_{\rm c}(Z)$ and $G=G_0(Z)$.
These quantities, $p_{\rm c}(Z)$ and $G_0(Z)$, are independent of $\alpha$ and identical for repulsive and cohesive systems within EMT.
Details of the estimation procedure are given in the Supplemental
Material~\cite{Supple}.

Figure~\ref{3figures2}(a) shows $p$ as a function of $Z-Z_{\rm iso}$
for both repulsive and cohesive particles, together with the marginal-stability line $p_{\rm c}(Z)$.
Here, $Z_{\rm iso}=6$ in three dimensions.
The figure represents a mechanical stability diagram, with the shaded region indicating EMT-stable states satisfying $p\le p_{\rm c}(Z)$.
The repulsive data collapse onto the marginal-stability line $p_{\rm c}(Z)$, independent of the loading protocol, indicating that $p$ is uniquely determined by $Z$, consistent with the marginal-stability condition.
In contrast, the cohesive data exhibit protocol dependence and systematically lie below this line within the EMT-stable region.
This demonstrates that $p$ cannot be expressed as a single-valued
function of $Z$, revealing the breakdown of marginal
stability, $p<p_{\rm c}(Z)$, in cohesive packings.
Note that data with negative pressure also appear below $p_{\rm c}(Z)$ in Fig.~\ref{3figures2}(a), implying that states with $p<0$ remain mechanically stable as long as the EMT stability condition $p\le p_{\rm c}(Z)$ is satisfied.

The breakdown of marginal stability is further supported by the
vDOS $D(\omega)$ for cohesive particles
at $\phi=0.60$ during compression, shown in the inset of Fig.~\ref{3figures2}(a), which compares the spectra obtained from the original Hessian $M$
[Eq.~\eqref{eq:M}] and the corresponding unstressed Hessian.
Near marginal stability, these spectra differ in the low-frequency
regime owing to excess soft modes \cite{Wyart2005a}.
In contrast, the two spectra shown in the inset nearly coincide,
indicating the suppression of these modes.
This provides additional evidence for the breakdown of marginal stability
in cohesive packings.

Figure~\ref{3figures2}(b) shows the shear modulus $G$ as a function of $Z-Z_{\rm iso}$, together with the bare shear modulus $G_0(Z)$ introduced in EMT.
For repulsive particles, the data follow $G=G_0(Z)$, consistent with the marginal-stability condition.
By contrast, cohesive particles systematically exhibit $G>G_0(Z)$, as expected from EMT for states with $p<p_{\rm c}(Z)$.
This excess rigidity reflects the breakdown of marginal stability identified in Fig.~\ref{3figures2}(a).

To quantify this relation, Eq.~\eqref{eq:G} yields a generalized scaling law:
\begin{align}
\frac{G-G_0(Z)}{G_0(Z)} =
\sqrt{\frac{p_{\rm c}(Z)-p}{p_{\rm c}(Z)}},
\label{eq:general_scaling}
\end{align}
which connects the excess rigidity $(G-G_0)/G_0$ to the breakdown of marginal stability $(p_{\rm c}-p)/p_{\rm c}$.
Figure~\ref{3figures2}(c) shows the corresponding scaling plot, where cohesive data collapse onto the EMT prediction (solid line) without fitting parameters, except for a few points at small packing fraction $\phi$.
The generalized scaling law is also verified for various values of $\alpha$ obtained using pressure-controlled protocols~\cite{Supple}.
Because both $p_{\rm c}(Z)$ and $G_0(Z)$ are determined from the repulsive ($\alpha=0$) data, this collapse provides a parameter-free test of EMT for cohesive packings.
The deviation at small $\phi$ likely reflects the breakdown of the mean-field assumption underlying EMT, possibly due to spatial heterogeneity in low-density configurations.
This quantitative agreement directly verifies the EMT prediction, Eq.~\eqref{eq:G}, and indicates that the hysteresis of $G(p)$ shown in Fig.~\ref{P_G}(c) arises because $G$ is no longer uniquely determined by $p$, with the coordination number $Z$ retaining memory of the compaction history for cohesive particles.

{\it Discussion and conclusion.---}
We numerically and theoretically demonstrate that cohesive interactions
induce pronounced mechanical hysteresis in jammed granular matter,
even when the interparticle force law is free from microscopic hysteresis.
Our results, supported by EMT, indicate that this history dependence arises from a generic breakdown of marginal stability,
where cohesive packings systematically violate the unique relation between
pressure and coordination number that holds in purely repulsive systems.
This highlights the history-dependent nature of the shear modulus.

For artificial Hessian ensembles constructed for repulsive packings, a related scaling relation~\cite{DeGiuli2014a} and the breakdown of marginal stability~\cite{Brito2018} were previously verified.
Our results extend this picture to physically realistic cohesive packings, where attractive interactions naturally induce a departure from marginal stability.
However, the mechanism underlying cohesion-induced breakdown of marginal stability remains unclear.
In general, cohesive packings can host multiple stable configurations at a given pressure owing to coexisting repulsive and attractive contacts, weakening the unique pressure-coordination relation of repulsive systems.
Using a minimal cohesive model, we show that this mechanism is generic,
while additional effects such as friction, interaction hysteresis, and nonlinear
contact elasticity may further enrich the behavior.
The present findings demonstrate how attractive interactions fundamentally reshape the stability landscape near jamming.

\begin{acknowledgments}
  We thank H. Hayakawa, A. Ikeda, H. Ikeda, T. Kawasaki, and H. Yoshino
for fruitful discussions.
This work was supported by JST ERATO Grant No. JPMJER2401, JSPS KAKENHI under Grant Nos.
JP23K03248, JP25H01519, and JP25K01063.
\end{acknowledgments}

\bibliography{main}

\clearpage

\setcounter{equation}{0}
\setcounter{figure}{0}

\renewcommand{\theequation}{S\arabic{equation}}
\renewcommand{\thefigure}{S\arabic{figure}}

\begin{center}
\textbf{\Large Supplemental Material: }
\end{center}

\section{Details of numerical simulations}

In our simulations, we consider $N$ frictionless monodisperse particles
of mass $m$ in a cubic box of size $L$.
Particle dynamics are integrated using the SLLOD equations of motion
under Lees--Edwards boundary conditions:
\begin{align}
\frac{d\boldsymbol r_i}{dt} &= \dot\gamma(t)\,y_i\,\boldsymbol e_x
  + \frac{\boldsymbol p_i}{m}, \label{eq:r}\\
\frac{d\boldsymbol p_i}{dt} &=
-\dot\gamma(t)\,p_{i,y}\,\boldsymbol e_x
  + \sum_{j \neq i} \boldsymbol F_{ij}, \label{eq:p}
\end{align}
where $\boldsymbol r_i=(x_i,y_i,z_i)$ denotes the position of particle
$i$, $\dot\gamma(t)$ is the shear rate, and $\boldsymbol e_x$ is the unit
vector along the $x$ direction.
The vector
$\boldsymbol p_i = m\dot{\boldsymbol r}_i
- \dot\gamma(t)y_i\boldsymbol e_x$
denotes the peculiar momentum of particle $i$.
During compression and decompression, we set $\dot\gamma=0$, while a
small shear deformation is applied when measuring the shear modulus.

The force between particles $i$ and $j$ is given by
\begin{align}
\boldsymbol F_{ij}
=
\left\{
f(r_{ij}) + f^{\rm(d)}\left (r_{ij},v^{\rm(n)}_{ij} \right )
\right\}\boldsymbol n_{ij},
\end{align}
where $f(r)$ is the interaction force defined in
Eq.~\eqref{eq:force} of the main text.
Here, $r_{ij}=|\boldsymbol r_i-\boldsymbol r_j|$ is the interparticle
distance and
$\boldsymbol n_{ij}=(\boldsymbol r_i-\boldsymbol r_j)/r_{ij}$
is the unit normal vector pointing from particle $j$ to $i$.
The normal relative velocity is defined as
\begin{align}
v^{\rm(n)}_{ij} =
(\dot{\boldsymbol r}_i-\dot{\boldsymbol r}_j)
\cdot\boldsymbol n_{ij}.
\end{align}
The dissipative contact force is given by
\begin{align}
f^{\rm(d)}(r,v) = -\eta v\,\Theta(d-r),
\end{align}
where $\eta$ is the viscous coefficient and
$d=(1-\alpha)\ell_{\rm c}$ is the particle diameter.
The Heaviside function $\Theta(x)$ satisfies $\Theta(x)=1$ for $x>0$
and $\Theta(x)=0$ otherwise, so that the dissipative force acts only
when particles overlap.

During compression and decompression, the packing fraction $\phi$ is
varied stepwise with increment $\Delta\phi$.
At each step, the system size $L$ is changed while applying an affine
transformation to the particle configuration, after which the particle
positions are relaxed for a time $T_{\rm R}$.
The resulting mechanically stable configuration at each $\phi$ is then stored for
subsequent measurements.

The shear modulus $G$ at each state is measured by applying an
oscillatory shear deformation with shear rate
\begin{align}
\dot\gamma(t)=\gamma_0\Omega\cos(\Omega t)
\label{eq:Shear}
\end{align}
to the stored configuration, where $\gamma_0$ and $\Omega$ denote the
strain amplitude and the angular frequency, respectively.
The oscillatory shear is applied for four cycles, and the shear modulus
is obtained from the last cycle as the storage modulus
\begin{align}
G =
\frac{\Omega}{\pi}
\int_0^{2\pi/\Omega}
dt\,\frac{\sigma(t)\sin(\Omega t)}{\gamma_0}.
\end{align}
Here, $\sigma(t)$ denotes the shear stress given by
\begin{align}
\sigma =
-\frac{1}{L^3}
\sum_i
\left\{
\frac{p_{i,x}p_{i,y}}{m}
+
\sum_{j>i} x_{ij}F_{ij,y}
\right\},
\end{align}
where
$\boldsymbol r_{ij}=\boldsymbol r_i-\boldsymbol r_j
=(x_{ij},y_{ij},z_{ij})$.

After the shear measurement, the pressure $p$ and coordination
number $Z$ are evaluated.
The pressure is calculated as
\begin{align}
p =
\frac{1}{3L^3}
\sum_i
\left\{
\frac{|\boldsymbol p_i|^2}{m}
+
\sum_{j>i} \boldsymbol r_{ij} \cdot \boldsymbol F_{ij} 
\right\},
\end{align}
while the coordination number is defined as
\begin{align}
Z = \frac{2N_{\rm pair}}{N},
\end{align}
where $N_{\rm pair}$ is the number of interacting particle pairs with
$f(r_{ij})\neq0$.
Rattler particles with fewer than four interacting neighbors are excluded when
evaluating $Z$.

We use
$\eta/(k\tau)=1.0$,
$\phi_{\rm I}=0.01$,
$\phi_{\rm max}=0.68$,
$\Delta\phi=0.0001$,
and $T_{\rm R}/\tau=100$.
All simulations in the main text were performed with
$N=4000$ particles.
We verified that the shear modulus and the hysteresis behavior
remain quantitatively unchanged for a smaller system size
$N=1000$, indicating that the system size used in the main text is
sufficiently large.
The equations of motion are integrated using the leapfrog algorithm with time step $\Delta t/\tau=0.05$.

Mechanical equilibrium of the configuration is confirmed after the
relaxation step, where the total force acting on each
particle becomes negligibly small compared with the characteristic
interaction force scale. Increasing the relaxation time $T_{\rm R}$
does not change the measured observables, confirming that the system
has reached mechanical equilibrium.

The strain amplitude and angular frequency are set sufficiently
small,
$\gamma_0=10^{-4}$ and $\Omega\tau=10^{-4}$,
so that the response remains in the linear regime.
We verified that the measured shear modulus is independent of
$\gamma_0$ and $\Omega$ within this range.

In the numerical analysis, states with vanishing pressure are
identified as those satisfying
$|p|<p_{\rm tol}$,
where the numerical threshold is set to
$p_{\rm tol}/E=10^{-7}$.
We verified that the results remain unchanged for
$p_{\rm tol}/E=10^{-8}$.

For comparison with previous studies
\cite{Koeze2018,Koeze2020},
the interaction force can also be written as
\begin{align}
f(r) =
\begin{cases}
k(\sigma-r) & r < (1+a)\sigma, \\
-k\left[(1+2a)\sigma-r\right]
& (1+a)\sigma \le r < (1+2a)\sigma, \\
0 & r \ge (1+2a)\sigma ,
\end{cases}
\end{align}
where $\sigma$ denotes the distance where the force changes sign
and $a$ controls the strength of the attractive interaction.
In our notation, this length corresponds to
\begin{align}
\sigma = (1-2\alpha)\ell_{\rm c},
\end{align}
and the attraction parameters are related by
\begin{align}
a = \frac{\alpha}{1-2\alpha}.
\end{align}
Thus, the two formulations are equivalent up to a
trivial rescaling of parameters.

\section{Additional plots of the simulation data}

\begin{figure}[htbp]
  \begin{center}
       \includegraphics[width=0.8\linewidth]{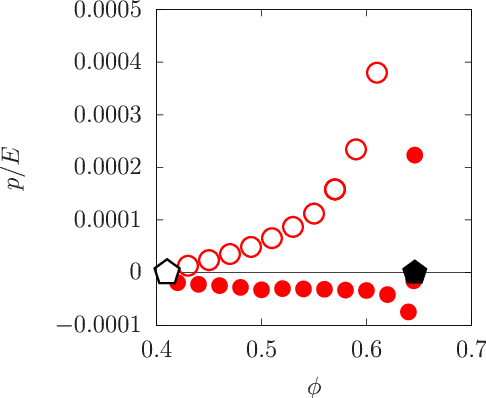}
  \end{center}
  \caption{
Enlarged view of the low-pressure regime of Fig.~\ref{P_G}(a),
showing the pressure $p$ as a function of the packing fraction
$\phi$.
Compression and decompression data are shown by open and closed
symbols, respectively.
The packing fraction $\phi_{p=0}$ where the pressure vanishes
is indicated by black pentagons.
}
    \label{phi_P_CP_Zoom}
\end{figure}

Figure~\ref{phi_P_CP_Zoom} shows an enlarged view of the
low-pressure regime of Fig.~\ref{P_G}(a), which clarifies the
behavior near $p\simeq0$ that is difficult to resolve in the
main figure.
The plot displays the pressure $p$ as a function of the packing
fraction $\phi$.
During compression (open symbols), $p$ starts to increase at
$\phi_{p=0}$, where the pressure vanishes.
During decompression (closed symbols), the pressure again vanishes
at $\phi_{p=0}$, while negative pressure persists for
$\phi < \phi_{p=0}$.

\begin{figure}[htbp]
  \begin{center}
       \includegraphics[width=0.8\linewidth]{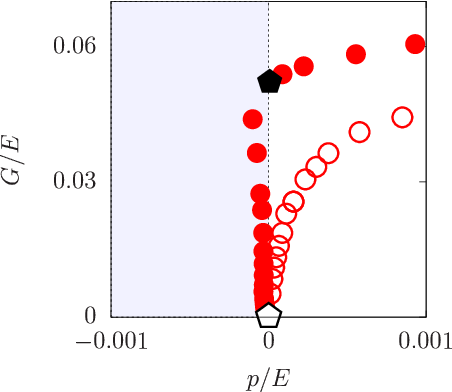}
  \end{center}
  \caption{
Shear modulus $G$ as a function of the pressure $p$ for cohesive
particles.
The region with negative pressure ($p<0$) is highlighted by the
blue shaded area.
}
    \label{G_P_CP_normal}
\end{figure}

Figure.~\ref{G_P_CP_normal} presents the shear modulus $G$ as a function
of the pressure $p$ for cohesive particles.
Clear hysteresis is observed between compression and decompression.
During decompression, a nonzero shear modulus persists even for
negative pressure ($p<0$).

\begin{figure}[htbp]
  \begin{center}
       \includegraphics[width=0.8\linewidth]{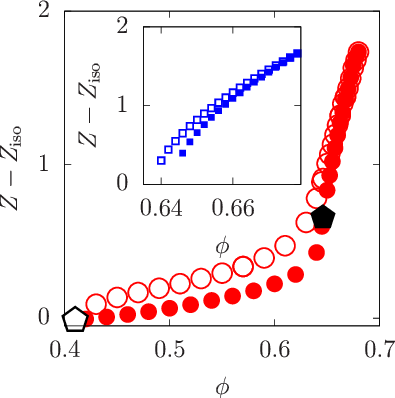}
  \end{center}
  \caption{
Excess coordination number $Z-Z_{\rm iso}$ as a function of the
packing fraction $\phi$ for cohesive particles (main panel) and
repulsive particles (inset).
}
    \label{Z_phi_CP}
\end{figure}

Figure~\ref{Z_phi_CP} presents the excess coordination number
$Z-Z_{\rm iso}$ as a function of the packing fraction $\phi$
for cohesive and repulsive particles.
For repulsive particles, $Z-Z_{\rm iso}$ exhibits hysteresis
between compression and decompression at a given $\phi$.
For cohesive particles, hysteresis is also observed; however,
during decompression, $Z-Z_{\rm iso}$ remains nonzero even for
$\phi < \phi_{p=0}$, indicating the persistence of contacts.
This behavior is consistent with that of the shear modulus $G$
shown in the main text.

\section{Pair distribution function}

\begin{figure}[htbp]
  \begin{center}
       \includegraphics[width=0.8\linewidth]{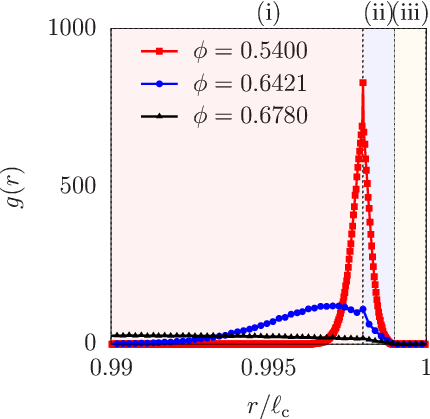}
  \end{center}
  \caption{
Pair distribution function $g(r)$ as a function of the
interparticle distance $r$ for cohesive particles with different
packing fractions $\phi$ during compression.
  The repulsive (i), stabilizing attractive (ii), and locally destabilizing
  regimes (iii) are indicated by shaded regions.
}
\label{Gr_a1e-3}
\end{figure}

Figure~\ref{Gr_a1e-3}, we plot the pair distribution function
$g(r)$ as a function of the interparticle distance $r$
for cohesive particles with different packing fractions $\phi$
during compression.
The repulsive, stabilizing attractive, and locally destabilizing
regimes are indicated by shaded regions.
We find that $g(r)>0$ in both the repulsive and stabilizing
attractive regimes, whereas $g(r)=0$ in the locally destabilizing
regime.
This indicates that contacts in the locally destabilizing regime
are absent in the mechanically relaxed configurations.
Near $\phi_{p=0}$ ($\phi=0.540$), a sharp peak appears in the
vicinity of the boundary between the repulsive and stabilizing
attractive regimes, where $f(r)=0$.
This peak reflects the accumulation of particle pairs near the
force-balance distance.

\begin{figure}[htbp]
  \begin{center}
       \includegraphics[width=0.8\linewidth]{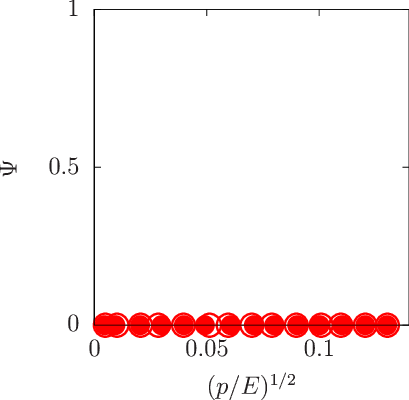}
  \end{center}
  \caption{
Fraction $\Psi$ of contacts in the locally destabilizing regime
($s_{ij}=-1$) as a function of the pressure $p$ for cohesive particles.
The data show $\Psi=0$ for all configurations.
}
\label{Psi_P}
\end{figure}

Here, we introduce the fraction $\Psi$ of contacts in the locally
destabilizing regime ($s_{ij}=-1$) for cohesive particles.
Figure~\ref{Psi_P} shows $\Psi$ as a function of the pressure $p$.
We find that $\Psi=0$ for all configurations obtained in our
simulations, indicating that contacts in the locally destabilizing regime are absent in mechanically stable states.

\section{Details of EMT}

In this section, we apply effective medium theory (EMT) to cohesive
particles following the procedure introduced in
Ref.~\cite{Mizuno2024}.

\subsection{Setup}

Within EMT, the particle system described by the Hessian matrix in
Eq.~\eqref{eq:M} is represented by a three-dimensional random spring
network. This disordered network is later replaced by an effective
homogeneous medium within a mean-field approximation. In this
construction, $N$ particles are placed on lattice sites with
coordination number $Z_0$, and neighboring sites are connected by
springs. Each bond $\langle ij\rangle$ represents a possible contact
in the original particle packing.

The displacement vector of particle $i$ is denoted by $\vec{u}_i$.
We introduce the $3N$-dimensional displacement vector
$|u\rangle=[\vec{u}_1,\ldots,\vec{u}_N]$.
Similarly, we define
$|F\rangle=[\vec{F}_1,\ldots,\vec{F}_N]$,
where $\vec{F}_i$ is the external force acting on particle $i$.
The elastic energy relative to the reference configuration is written
as
\begin{align}
\mathcal{U}
=
\frac{1}{2}\langle u | M_{\rm R} | u \rangle
-
\langle u | F \rangle .
\end{align}
Here, $M_{\rm R}$ denotes the Hessian matrix of the random spring
network,
\begin{equation}
M_{\rm R}
=
\sum_{\langle ij\rangle}
|ij\rangle
k_{ij}
\left[
s_{ij}\,\vec{n}_{ij}\otimes\vec{n}_{ij}
-
e\left(I-\vec{n}_{ij}\otimes\vec{n}_{ij}\right)
\right]
\langle ij|.
\label{eq:MR}
\end{equation}
The sum runs over $NZ_0/2$ pairs of connected particles. Here, $k_{ij}$
is the spring stiffness and $s_{ij}$ represents the local stability
sign induced by the cohesive interaction. The parameter $e$ denotes
the prestress, corresponding to $f_{ij}/(k r_{ij})$ in the
original particle system. Typically, $e$ is related to the pressure
$p$ as
\begin{align}
p \simeq \frac{3k\phi e}{\pi\sigma}.
\end{align}
In mechanical equilibrium, the displacement field satisfies
\begin{align}
|u\rangle = M_{\rm R}^{-1}|F\rangle ,
\end{align}
where $M_{\rm R}^{-1}$ plays the role of the Green function.
The randomness of the network is introduced through the probability
distributions of $k_{ij}$ and $s_{ij}$:
\begin{align}
P_k(k_{ij})
&=
\frac{Z}{Z_0}\delta(k_{ij}-k)
+
\left(1-\frac{Z}{Z_0}\right)\delta(k_{ij}),
\\
P_s(s_{ij})
&=
\Psi\delta(s_{ij}+1)
+
\left(1-\Psi\right)\delta(s_{ij}-1),
\end{align}
where $Z$ is the coordination number and $\Psi$ is the fraction of
contacts in the locally destabilizing regime introduced in the
original particle system.

\subsection{Mean-field description}

Within EMT, the disordered network is replaced by an effective
homogeneous medium. The corresponding Hessian matrix is written as
\begin{equation}
M_{\rm eff}
=
\sum_{\langle ij\rangle} |ij\rangle
\left[
k^{\parallel}\vec{n}_{ij}\otimes\vec{n}_{ij}
-
e k^{\perp}\left(I-\vec{n}_{ij}\otimes\vec{n}_{ij}\right)
\right]
\langle ij|.
\label{eq:Me}
\end{equation}
Here, $k^{\parallel}$ and $k^{\perp}$ denote the effective longitudinal
and transverse stiffnesses of the medium, respectively.
These parameters are determined so that the averaged Green function of
the random spring network coincides with that of the effective medium,
\begin{equation}
\langle M_{\rm R}^{-1} \rangle = M_{\rm eff}^{-1},
  \label{M:av}
\end{equation}
where $\langle\cdot\rangle$ denotes the average over the probability
distributions of $k_{ij}$ and $s_{ij}$.

The elastic response of the effective medium is governed by the two
parameters $k^{\parallel}$ and $k^{\perp}$.
Following Ref.~\cite{Mizuno2024}, the shear modulus of the effective
medium (and hence of the random network) is written as
\begin{equation}
G = \beta E \left(\frac{k^{\parallel}}{k}-2e \frac{k^{\perp}}{k}\right),
  \label{G:MR}
\end{equation}
where $\beta$ is a dimensionless geometric constant.
Here, $E$ denotes the stress scale introduced in the main text.
The effective stiffnesses depend on $Z$, $e$, and $\Psi$, which
characterize the random network, leading to the corresponding
parameter dependence of $G$.
The effective stiffnesses $k^{\parallel}$ and $k^{\perp}$ are determined below from the EMT self-consistency condition.

\subsection{Determination of effective stiffnesses}

For notational simplicity, we introduce the Green functions of the
random network and the effective medium as
\begin{equation}
\mathcal{G}_{\rm R}=M_{\rm R}^{-1}, \qquad
\mathcal{G}_{\rm eff}=M_{\rm eff}^{-1}.
\end{equation}
The EMT condition in Eq.~\eqref{M:av} can then be written as
\begin{equation}
\langle \mathcal{G}_{\rm R} \rangle = \mathcal{G}_{\rm eff}.
\label{G:av}
\end{equation}

Following the standard EMT procedure~\cite{Mizuno2024},
we express the Green function of the random network as
\begin{equation}
\mathcal{G}_{\rm R}
=
\mathcal{G}_{\rm eff}
+
\mathcal{G}_{\rm eff} T \mathcal{G}_{\rm eff}.
\label{G:T}
\end{equation}
Here, $T$ is the transfer matrix, which can be expanded as
\begin{equation}
T
=
\sum_{\langle ij\rangle} T_{\langle ij\rangle}
+
\sum_{\langle ij\rangle}
\sum_{\langle ln\rangle\neq\langle ij\rangle}
T_{\langle ij\rangle}\mathcal{G}_{\rm eff}T_{\langle ln\rangle}
+\cdots .
\end{equation}
The single-bond transfer matrix is given by
\begin{align}
T_{\langle ij\rangle} =
|ij\rangle
\Bigg[
&\frac{k^\parallel - s_{ij}k_{ij}}
{1-(k^\parallel-s_{ij}k_{ij})g}
\,\vec{n}_{ij}\otimes\vec{n}_{ij}
\nonumber \\
&-
\frac{e(k^\perp-k_{ij})}
{1+e(k^\perp-k_{ij})g}
\left(I-\vec{n}_{ij}\otimes\vec{n}_{ij}\right)
\Bigg]
\langle ij| ,
\end{align}
where $g$ denotes the relevant scalar component of the effective
Green function $\mathcal{G}_{\rm eff}$.
For an isotropic effective medium, the trace of the Green function
yields the scalar quantity $g$, which is related to the effective
stiffnesses as
\begin{equation}
g = \frac{6}{Z_0}\frac{1}{k^\parallel-2ek^\perp}.
\label{g1}
\end{equation}

By averaging both sides of Eq.~\eqref{G:T} and using
Eq.~\eqref{G:av}, we obtain the EMT self-consistency condition
\begin{equation}
\langle T\rangle = 0 .
\end{equation}
Considering only single-bond scattering processes yields
\begin{equation}
g =
-\frac{k^\perp-k(Z/Z_0)}
{ek^\perp(k^\perp-k)}
\label{g2}
\end{equation}
and
\begin{align}
&\left [1-(k^\parallel+k)g \right ]
\left [k^\parallel-k(Z/Z_0)-k^\parallel(k^\parallel-k)g \right ]
\nonumber\\
&=
2(Z/Z_0)k\left (k^\parallel g-1 \right )\Psi .
\label{g3}
\end{align}
Equations~\eqref{g1}--\eqref{g3} determine the effective stiffnesses.

\subsection{Asymptotic analysis}

We introduce $\delta Z = Z - Z_{\rm iso}$ and assume $\delta Z \ll 1$
to obtain an asymptotic solution of Eqs.~\eqref{g1}--\eqref{g3}.
Following Ref.~\cite{Mizuno2024}, we expand the effective stiffnesses as
\begin{align}
k^\parallel &= k_1^\parallel \delta Z + O(\delta Z^2), \\
k^\perp &= k_0^\perp + O(\delta Z),
\end{align}
while the prestress parameter satisfies $e = O(\delta Z^2)$.
Using these expansions together with Eqs.~\eqref{g1}--\eqref{g3},
we perturbatively determine $k_1^\parallel$ and $k_0^\perp$.
Substituting the resulting expressions into Eq.~\eqref{G:MR} and
retaining the leading order in $\delta Z$, we obtain
\begin{align}
G =
\frac{E}{2(Z_0-6)(1-2\Psi)}
\,
\delta Z
\left(
1+\sqrt{\frac{e_{\rm c}-e}{e_{\rm c}}}
\right),
\end{align}
where
\begin{align}
e_{\rm c} =
\frac{Z_0}{144(Z_0-6)}
\frac{\delta Z^2}{1-2\Psi}.
\end{align}

In the present system, we numerically find $\Psi=0$, as shown in
Fig.~\ref{Psi_P}.
In this case, the above expression reduces to that obtained for
repulsive particles in Ref.~\cite{Mizuno2024}.
Because $e$ is proportional to the pressure $p$, the factor
$(e_{\rm c}-e)/e_{\rm c}$ can be rewritten as
$(p_{\rm c}-p)/p_{\rm c}$, where the critical pressure is given by
$p_{\rm c}=\Pi_0 \delta Z^2$ with a constant $\Pi_0$.
Finally, defining $E_0 = E/[2(Z_0-6)]$, we obtain the theoretical
result given in Eq.~\eqref{eq:G} of the main text.

\section{Estimation of $p_{\rm c}$ and $G_0$}

To estimate the critical pressure $p_{\rm c}(Z)$ and the bare shear
modulus $G_0(Z)$, we use the relations
\begin{align}
p = p_{\rm c}(Z) = \Pi_0 (Z-Z_{\rm iso})^2, \\
G = G_0(Z) = E_0 (Z-Z_{\rm iso}),
\end{align}
which follow from Eqs.~\eqref{eq:G}--\eqref{eq:pc} under the marginal
stability condition for purely repulsive particles.
Using the data for repulsive particles during compression with
$Z-Z_{\rm iso}<1$, we estimate the coefficients
$\Pi_0/E = 0.0067$ and $E_0/E = 0.0435$ by fitting the relations above.
The functions $p_{\rm c}(Z)=\Pi_0 (Z-Z_{\rm iso})^2$ and
$G_0(Z)=E_0 (Z-Z_{\rm iso})$ with these parameters are then used in the
analysis of the simulation data.

\section{Vibrational density of states}

In this section, we examine the vibrational density of states $D(\omega)$ (vDOS) calculated from the eigenvalues ${\lambda_n}$ of the Hessian $M$ [Eq.~\eqref{eq:M}] by computing the corresponding frequencies $\omega_n = \sqrt{\lambda_n/m}$ and constructing a histogram of $\omega_n$, normalized such that $\int D(\omega) d\omega = 1$.
For comparison, we also compute the vDOS using the
unstressed Hessian $M^{\rm (un)}$, defined by removing the
stress term from Eq.~\eqref{eq:M}:
\begin{equation}
M^{\rm (un)}=
\sum_{\langle ij\rangle}
|ij\rangle
   s_{ij}k\, \vec{n}_{ij}\otimes\vec{n}_{ij}
\langle ij|.
\label{eq:Mun}
\end{equation}

For repulsive particles, it is well known that the vDOS obtained from
$M^{\rm (un)}$ exhibits a plateau above a characteristic frequency
$\omega^*$, which depends on the distance from the jamming point \cite{Wyart2005a}.
When the original Hessian $M$ is used, the plateau persists but a tail
appears below $\omega^*$ due to marginal stability.
As a result, a clear difference between the two spectra emerges in the
low-frequency regime $\omega < \omega^*$.

\begin{figure}[htbp]
  \begin{center}
       \includegraphics[width=0.8\linewidth]{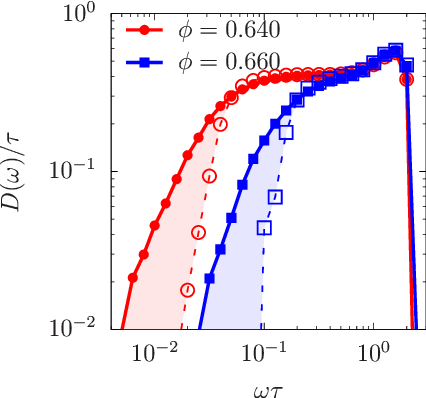}
  \end{center}
  \caption{
Vibrational density of states $D(\omega)$ for repulsive particles
 during compression with different packing fractions $\phi$.
Results obtained from the original Hessian $M$ are shown by closed symbols, while those obtained from the unstressed Hessian
$M^{\rm (un)}$ are shown by open symbols.
The shaded regions highlight the excess low-frequency modes
associated with marginal stability.
}
\label{DOS_a0}
\end{figure}

This behavior is confirmed in Fig.~\ref{DOS_a0}, which shows
$D(\omega)$ for repulsive particles with different packing fractions $\phi$ during decompression.
The spectra obtained from $M$ (closed symbols) exhibit an excess of
low-frequency modes compared with those obtained from
$M^{\rm (un)}$ (open symbols), as highlighted by the shaded regions.

\begin{figure}[htbp]
  \begin{center}
       \includegraphics[width=0.8\linewidth]{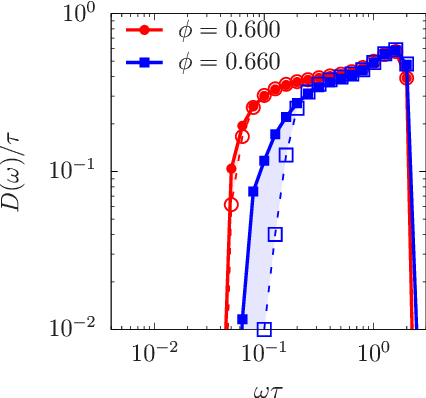}
  \end{center}
  \caption{
Vibrational density of states $D(\omega)$ for cohesive particles during compression with
different packing fractions $\phi$.
Closed and open symbols correspond to the spectra obtained from
$M$ and $M^{\rm (un)}$, respectively.
The reduced difference between the two spectra in the low-frequency
regime indicates the breakdown of marginal stability.
}
\label{DOS_a1e-3}
\end{figure}

The situation changes for cohesive particles, as shown in
Fig.~\ref{DOS_a1e-3}.
While the plateau structure remains visible, the difference between
the spectra obtained from $M$ and $M^{\rm (un)}$ becomes significantly smaller
in the low-frequency regime.
In particular, the excess low-frequency modes observed for repulsive
packings are strongly suppressed.
This reduction indicates that the marginal-stability condition is no
longer satisfied in cohesive packings.

\begin{figure}[htbp]
  \begin{center}
       \includegraphics[width=0.8\linewidth]{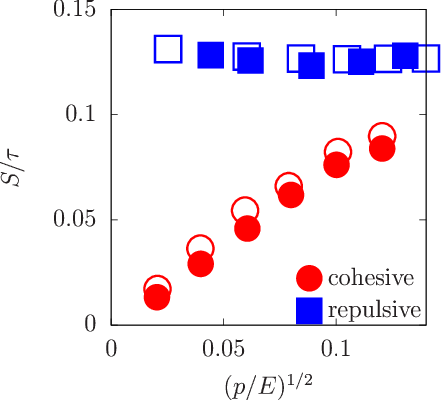}
  \end{center}
  \caption{
Area $S$ measuring the difference between the vibrational spectra obtained from the original Hessian $M$ and the unstressed Hessian $M^{\rm (un)}$, plotted as a function of $p^{1/2}$ for repulsive and cohesive particles during compression and decompression.
}
\label{S_P}
\end{figure}

To quantify the difference between the two spectra, we introduce the
area
\begin{align}
  S = \int_{-\infty}^\infty d(\log \omega  \tau)\,\Delta D(\omega),
\end{align}
where $\Delta D(\omega)$ denotes the difference between the spectra
obtained from $M$ and $M^{\rm (un)}$.
Figure~\ref{S_P} shows $S$ as a function of $p^{1/2}$ for repulsive
and cohesive particles during compression and decompression.
For repulsive particles, $S$ approaches a finite value as $p \to 0$,
reflecting the persistent excess of low-frequency modes associated
with marginal stability.
By contrast, for cohesive particles, $S$ decreases approximately
proportional to $p^{1/2}$ at small pressures, indicating that
$S \to 0$ as $p \to 0$.
This behavior indicates that the difference between the two spectra
vanishes in the jamming limit, providing further evidence for the
breakdown of marginal stability in cohesive packings.

\begin{figure}[htbp]
\centering
\includegraphics[width=0.8\linewidth]{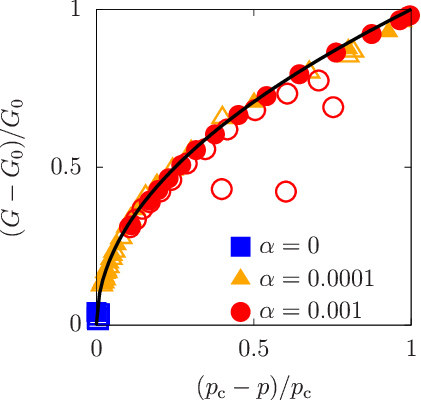}
\caption{
Scaling plot of the excess rigidity $(G-G_0)/G_0$ obtained from
pressure-controlled simulations as a function of the deviation from
marginal stability $(p_{\rm c}-p)/p_{\rm c}$ for
$\alpha=0.0$, $0.0001$, and $0.001$.
The solid line represents the EMT prediction,
Eq.~\eqref{eq:general_scaling} in the main text.
}
\label{Scale_G_CPL}
\end{figure}

\section{Scaling plot from pressure-controlled simulations with different $\alpha$}

To confirm that the generalized scaling relation is independent of
both the simulation protocol and the attraction strength $\alpha$,
we performed additional simulations using a pressure-controlled
protocol and several values of $\alpha$.
Instead of Eqs. \eqref{eq:r} and \eqref{eq:p}, the time
evolution is given by
\begin{align}
\frac{d\boldsymbol r_i}{dt} &= \dot\gamma(t)\,y_i\,\boldsymbol e_x
+ \dot\epsilon(t)\,\boldsymbol r_i
+ \frac{\boldsymbol p_i}{m}, \\
\frac{d\boldsymbol p_i}{dt} &=
-\dot\gamma(t)\,p_{i,y}\,\boldsymbol e_x
-\dot\epsilon(t)\,\boldsymbol p_i
+ \sum_{j \neq i} \boldsymbol F_{ij}, \\
\frac{dL}{dt} &= \dot \epsilon(t) L,
\end{align}
with the compressive strain rate
\begin{align}
\dot\epsilon(t) = (p_{\rm target} - p)/A ,
\end{align}
where $p_{\rm target}$ is the target pressure and $A$ is a constant.

Starting from the initial state with packing fraction $\phi_{\rm I}$,
the target pressure $p_{\rm target}$ is increased stepwise by $\Delta p$
until it reaches a maximum value $p_{\rm max}$; we refer to this
process as compression.
Subsequently, $p_{\rm target}$ is decreased stepwise by $\Delta p$,
which we refer to as decompression.
At each $p_{\rm target}$ during both compression and decompression,
the shear modulus $G$ and coordination number $Z$ are measured after
the system reaches mechanical equilibrium for a time $T_{\rm P}$.
The procedure is otherwise identical to the $\phi$-controlled
simulations used in the main text.

During the measurement process, we confirm that the measured pressure
$p$ coincides with $p_{\rm target}$.
Therefore, in the following figure, we use $p$ instead of
$p_{\rm target}$.
The numerical parameters are $\Delta p/E=10^{-5}$,
$p_{\rm max}/E=0.017$, $T_{\rm P}/\tau=400$, and
$A/(E\tau)=10$.

Figure~\ref{Scale_G_CPL} shows the scaling plot of $(G-G_0)/G_0$
as a function of $(p_c-p)/p_c$ obtained from the pressure-controlled
simulations for $\alpha=0.0$, $0.0001$, and $0.001$.
The data for different $\alpha$ and loading processes collapse onto
the EMT prediction in the same manner as in the
packing-fraction-controlled simulations shown in the main text,
although the data for the repulsive particles ($\alpha=0$) fall at the
origin of the plot.
These results demonstrate that the generalized scaling and the
associated breakdown of marginal stability are robust with respect
to both the simulation protocol and the interaction strength.
\end{document}